\documentclass[journal]{IEEEtran}
\ifCLASSINFOpdf
   \usepackage[pdftex]{graphicx}
\else
\fi
\usepackage{amsmath}
\usepackage{gensymb}

\usepackage{soul, color}

\begin{document}
\title{The Metaverse from a Multimedia Communications Perspective}

\author{Haiwei~Dong,~\IEEEmembership{Senior~Member,~IEEE,}
        % John~Doe,~\IEEEmembership{Fellow,~OSA,}
         and~Jeannie S.A. Lee,~\IEEEmembership{Senior~Member,~IEEE}% <-this % stops a space
% \thanks{M. Shell was with the Department
% of Electrical and Computer Engineering, Georgia Institute of Technology, Atlanta,
% GA, 30332 USA e-mail: (see http://www.michaelshell.org/contact.html).}% <-this % stops a space
% \thanks{J. Doe and J. Doe are with Anonymous University.}% <-this % stops a space
% \thanks{Manuscript received April 19, 2005; revised August 26, 2015.}
}

\maketitle

\begin{abstract}
% The abstract goes here.
eXtended reality (XR) technologies such as virtual reality and 360\textdegree{} stereoscopic streaming enable the concept of the Metaverse, an immersive virtual space for collaboration and interaction. To ensure high fidelity display of immersive media, the bandwidth, latency and network traffic patterns will need to be considered to ensure a user's Quality of Experience (QoE). In this article, examples and calculations are explored to demonstrate the requirements of the abovementioned parameters. Additionally, future methods such as network-awareness using reinforcement learning (RL) and XR content awareness using spatial or temporal difference in the frames could be explored from a multimedia communications perspective. 
\end{abstract}

% \begin{IEEEkeywords}
% IEEE, IEEEtran, journal, \LaTeX, paper, template.
% \end{IEEEkeywords}

% \fi
\IEEEpeerreviewmaketitle

\section{The Metaverse}
No longer hypothetical, the Metaverse is the concept of a fully immersive and universal virtual space for multi-user interaction, collaboration, and socializing, considered the next evolution of the Internet. The word is a portmanteau of the words ``meta" and ``universe", first seen in the science-fiction novel \textit{}{Snow Crash} in 1992. It describes a virtual world accessed via wearing a headset, an alternate realm where the streets are neon-lit, and each person is represented by a customized avatar.

The Metaverse depends on the convergence of multiple broad technologies that enable eXtended Reality (XR), which is an umbrella term for technologies that lie on the reality-virtuality continuum, namely virtual reality (VR), augmented reality (AR) and mixed reality (MR). Companies such as Microsoft and Facebook (now rebranded to Meta) have heavily invested in these technologies, and also envision a new economy where digital assets can be purchased or traded, and virtual land bought and sold.

For the display of virtual environments that comprise the Metaverse, this can be achieved via the development of a vector 3D environment involving virtual objects, scenes, and avatars. Alternately, a commonly-used method to transmit immersive metaverse content to the headset would be 360-degree stereoscopic video streaming. One such scenario would be experiencing a virtual walk through a futuristic cityscape, being able to choose different viewing perspectives and interact with different objects or other avatars that are present.

High fidelity display and interaction with such immersive media is of importance with regard to users' quality of experience (QoE). In addition to multi-view 360\textdegree{} video, 3 Degrees of Freedom (3DoF) head tracking involving the movements of yaw, pitch and roll and 6DoF inclusive of x, y, and z axes will need to be supported. Correspondence in the visual, vestibular and somatosensory systems is also required, to reduce cybersickness during head-mounted display (HMD) based XR.

Therefore, the resolution required is 4K, with multiple-view 360\textdegree{} videos with texture and depth, and minimal latency in response to human movement in the virtual environment. This requires a large amount of bandwidth, processing and low latency.

% In this article, we present insights into the fundamental multimedia streaming and communications for XR technologies that support the Metaverse, as well as the key challenges and promising solutions.

% For the display of virtual environments in the metaverse, this can be achieved via the development of a vector 3D environment involving virtual objects, scenes, and avatars. Another commonly-used method would be 360-degree pixel streaming of immersive content to the headset. In this article, we present an insight into the fundamental multimedia streaming and communications for the XR technologies that support the Metaverse, as well as the key challenges and solutions.
%\hl{to further define the scope and terminology}

\section{Extended Reality (XR) Communications}
Currently, the rendering of the XR content happens in the 
XR headset. Whereas the current computing power of mobile GPUs in the headset for XR is becoming a bottleneck with the rapidly evolving needs in the immersive experience. Given that Moore's Law is currently approaching its physical limit (transistor size is rapidly approaching the atomic level), it is believed that the future XR will be rendered in cloud servers. The XR content and related synchronization as well as measurements would be transmitted in the network between XR cloud and XR headsets. To enable XR streaming, it is important to quantify the amount of bandwidth, latency, and frame update rates that are needed. These requirements always closely relate to the physical characteristics of human perception.

\subsection{Low Latency Requirement}
% In interconnected virtual reality research conducted by MIT and Huawei France gives how much bandwidth, how little latency, and how high frame update rates are needed in VR transmission. The end-to-end latency is defined as less than 13ms which is based on the threshold value that human can perceive. The latency requirement comes from the physical characteristics of the human eyes, i.e., how much latency human eyes can detect. 

The end-to-end latency \cite{Untethered_VR} can be modeled as the sum of sensing time, rendering time, streaming time, and displaying time. The ideal end-to-end delay should be 7 ms because the vestibulo-ocular reflex (VOR) process takes 7 ms \cite{VR_AR_5G}. Here, the sensing time is for the headset and controller sensors to take measurement, mainly for localization and controller inputs, which is typically 400 microseconds. The rendering time is for both foreground and background which is around 5 to 11 ms depending on the complexity of the constructed virtual environment. Displaying time is to show the XR content in front of the human eye. If the framerate has to be 90 frames/s, the end-to-end latency has to be less than $1 \div 90=11.1$ ms. For the case of 120 frames/s, the end-to-end latency needs to be even less than 8.3 ms. For both cases, the rendering and displaying time have to be shortened as much as possible by an Application-Specific Integrated Circuit for saving more time in transmission, and thus leaving more room for the streaming time.

% \begin{equation}
%     L_{e2e}=T_{sensing}+T_{rendering}+T_{streaming}+T_{displaying}
% \end{equation}
% where $T_{sensing}$ is the time for the helmet and controller sensors to make measurement, mainly for localization and controller inputs, which is typically 400 microseconds. $T_{rendering}$ is the rendering time for both foreground and background which is around 5 to 11 milliseconds depending the complexity of the constructed virtual environment. $T_{displaying}$ is the time for displaying the XR contents in front of the human eyes. If we need the framerate to be 90frames/second, $L_{e2e}$ has to be less than $1 \div 90=11.1$milliseconds. For 120 frams/second case, $L_{e2e}$ needs to be even less than 8.3milliseconds. For both cases, the rendering and displaying time have to be shortened as much as possible by ASICs (Application-Specific Integrated Circuit) for saving more time in transmission (leaving more room for $T_{streaming}$).

\subsection{High Bandwidth Requirement}
% For bandwidth requirement, it is also calculated according to the characters of human eyes. That is, the human eye can distinguish 200 points in 1 degree. By using this as the resolution basis, the following figure shows how high the resolution is possible for different screen sizes and the distance from the eyes.

The main parameter to determine the XR content quality includes PPD (Pixels Per Degree), color depth, and frame refresh rate.
\begin{itemize}
    \item PPD is related to the display's resolution indicating the pixel number per degree. If the user's observation region is defined as W degrees in width and H degrees in height, the total pixel number for a VR monocular display is $(H \times PPD) \times (W \times PPD)$ \cite{360_degree_VR}. It is usually considered that the PPD of human eyes can distinguish is 60. In VR scenarios, particularly in 360 degrees panoramic VR, PPD is always set as 64. Legacy PPD metric can be 11, 21, and 32 as well.
    \item Color depth indicates how many bits to distinguish a color's grayscale. The combination of the three primary colors (i.e., red, green, and blue) is typically used to denote a color where the bit number of the color's depth is $2^{3x}$. In XR fields, the color depth is typically set as 8 bits to 12 bits.
    \item Frame refresh rate is defined as the number of refreshing frames/s. A higher frame refresh rate can effectively mitigate the issues of motion blur. The motion resolution that the human eyes can distinguish is up to 150 frames/s \cite{Interconnected_VR}. Typically, the frame refresh rate in XR is set as either 90 frames/s or 120 frames/s. Legacy frame refresh rate can be 30 or 60 frames/s.
\end{itemize}

Take HTC Vive or Oculus Rift as an XR device for example, its frame refresh rate is 90 frames/s, and resolution is 2K (i.e., 2160 $\times$ 1200). The downlink raw data rate can be calculated as $3 \times 8 \times 90 \times 2160 \times 1200 = 5,598,720,000$ bits/s which is around 5.6 Gb/s. As this data rate is very big, encoding codecs are used to make compression before transmission, such as H.264, H.265, or H.266. The compression rate of the above-mentioned three codecs is 102:1, 215:1, and 350:1, respectively. Thus, if H.264 is used to compress the 5.6 Gb/s raw data rate, the compressed data rate is $5.6 Gb/s \div 102=55$ Mb/s. Furthermore, the binocular compressed data rate is 110 Mb/s. For some legacy XR devices whose frame refresh rate is 30 frames/s, the compressed data rate for the monocular case is around 37 Mb/s.

If we consider another specific scenario, 360 degree VR, as an example, let's recalculate its data rate. The field of vision here is $360 \degree \times 180 \degree$ for panoramic view and $120\degree \times 120\degree$ for partial view. Even if the PPD is considered as 11, for panoramic VR scenarios, its raw data rate is $3 \times 360 \times 11 \times 180 \times 11 \times 8 \times 30=5.6$ Gb/s. For binocular VR displays, the raw data rate is 11.2 Gb/s. If 64 PPD is used in the calculation, the raw data rate can reach 2.3 Tb/s.

\subsection{Network Traffic Pattern}
\label{Section:Network Traffic Pattern}
The traffic in the network includes the downlink traffic (from XR cloud to XR headset) and uplink traffic (from XR headset to XR cloud) which have special characteristics, respectively.
\subsubsection{Downlink Data Flow}
The downlink data flow includes two types of packets. One comes from the rendered XR frames. As the packet size in practice has restrictions in MTU (maximum transmission unit), each XR frame for the left/right eye is divided into multiple packets (thousand bytes scale for each packet) transmitting almost at the same time leading to a burst. For example, if the framerate of the XR device is 60 frames/s, there would be two bursts of packets corresponding to the left and right eye at every $1\div60=16.7$ ms. Other packets include regularly transmitted synchronization packets which are around a few bytes at every 16.7 ms for the previous example. 

\subsubsection{Uplink Data Flow}
The uplink data flow is typically composed of the pose of the XR headset, sensory observation and control signals from hand controllers, and synchronization signals. The pose measurement includes position and orientation of the XR headset in the air. It is always described by a 6 DoF (degrees of freedom) which relies on inertial navigation. The uplink data frequency can be very high whereas the size of the uplink data packet is in the scale of a few Bytes to 400 Bytes.

\subsection{Quality of Experience (QoE)}
\label{Section:QoE}
\begin{figure}
    \centering
    \includegraphics[width=0.9\linewidth]{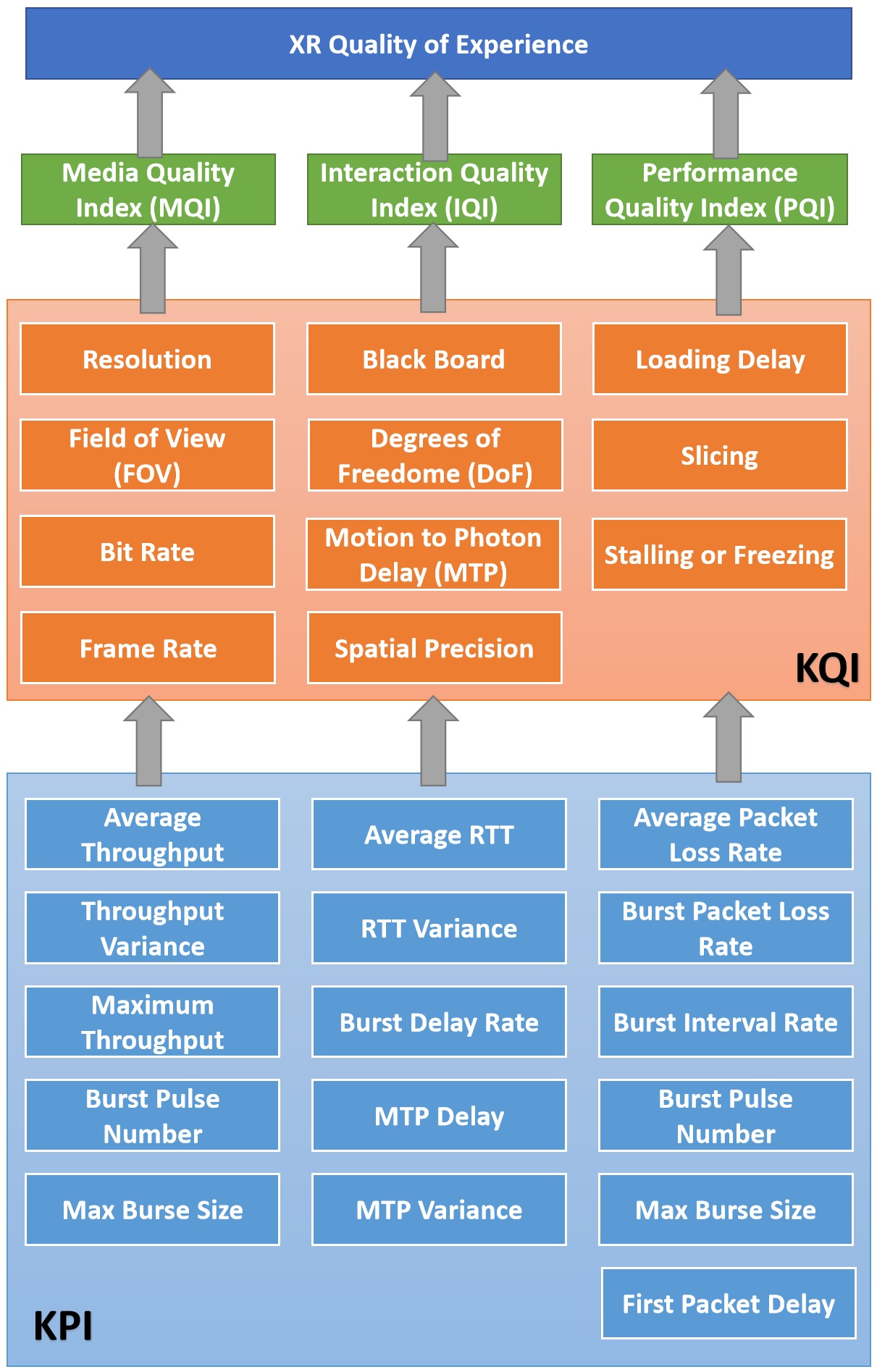}
    \caption{Factors to determine the quality of experience of users in XR scenarios.}
    \label{fig:XR_QoE}
\end{figure}
\begin{figure*}
    \centering
    \includegraphics[width=0.8\linewidth]{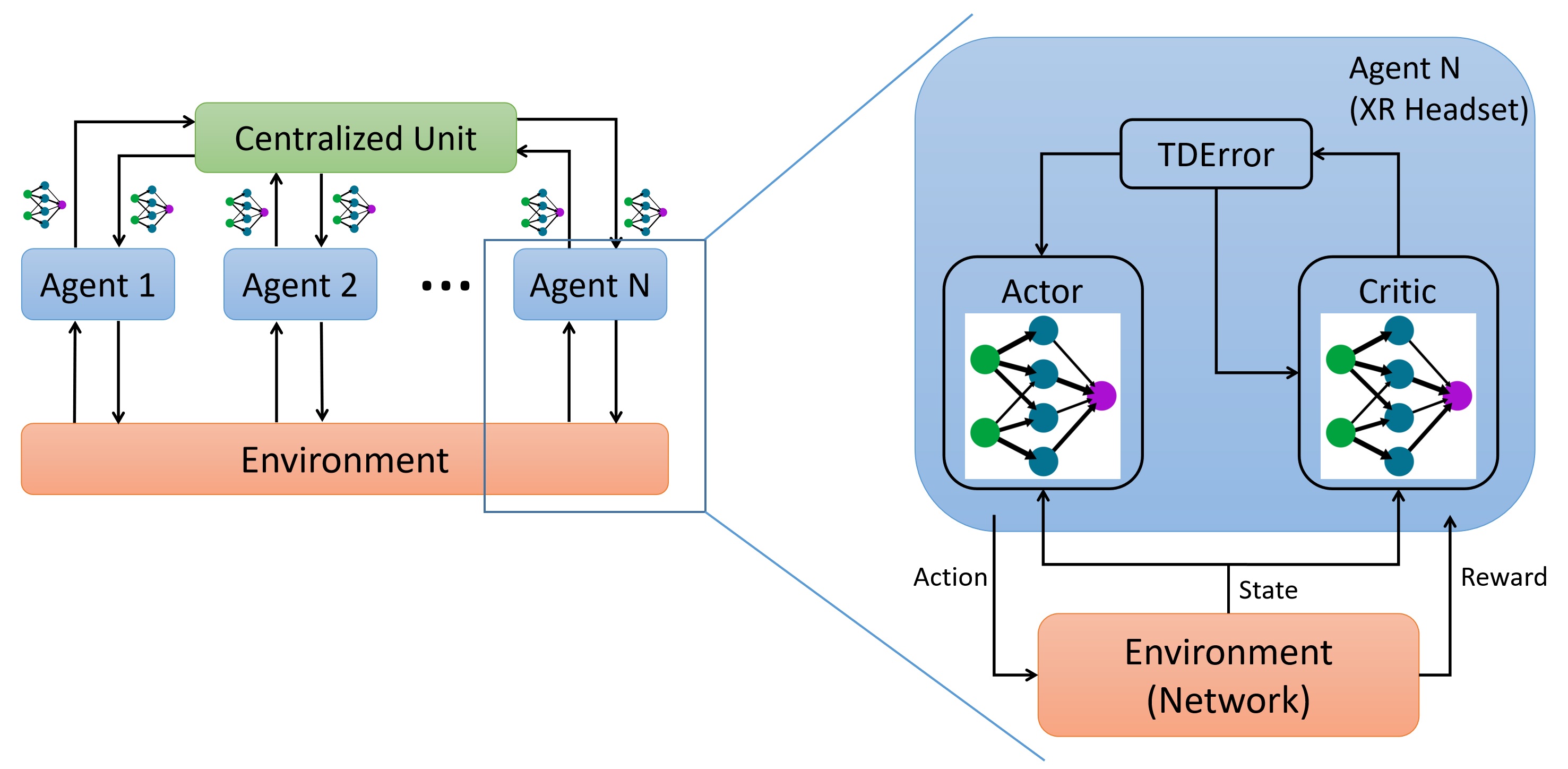}
    \caption{Centralized multi-agent reinforcement learning architecture to enable XR device to optimize its QoE by utilizing the network measurement.}
    \label{fig:MARL}
\end{figure*}
The quality of experience is a measure to quantify the overall users' subjective satisfaction with a service/device. There have been related standards in XR QoE including:
\begin{itemize}
    \item ITU-T G.QOE-VR standard entitled ``Influencing factors on quality of experience (QoE) for virtual reality (VR) services" proposed by International Telecommunication Union.
    \item 3GPP TR 26.929 standard entitled ``QoE parameters and metrics relevant to the Virtual Reality (VR) user experience" by 3rd Generation Partnership Project.
    \item ITU P1203 standard entitled ``Parametric bitstream-based quality assessment of progressive download and adaptive audiovisual streaming services over reliable transport" by International Telecommunication Union.
\end{itemize}

These standards typically combine three indexes, MQI (media quality index), IQI (interaction quality index), and PQI (performance quality index), to infer the XR QoE \cite{Huawei_Whitepaper}. Specifically, each index involves factors refining from KPI (key performance indexes) to KQI (key quality indexes) as shown in Fig.\ref{fig:XR_QoE}. The KPIs evaluate the performance from the engineering system perspective, whereas the KQIs and QoE are designed to quantify the feeling from human's perspective. These factors are usually summed together with different weights which are determined by inference (e.g., fuzzy inference \cite{QoE_AR}) from questionnaire results of human studies.

There are two ways to collect the data for the mentioned inference. One way is corresponding to the case when obtaining the XR data in the application layer from end-hosts or cloud servers. This is suitable for measuring the users' real experience which needs deployment of SDK in end-hosts or cloud servers. The other way indirectly measures the user's QoE by capturing and using the data from carriers' network pipe which needs the network probe such as in-band and out-of-the-band network telemetry.

% \subsection{Metric Outlook}

\section{XR - Network Integration}

\subsection{Network-Aware XR System}
The network connecting XR headsets and XR cloud (either located in campus or data center) has varying conditions due to concurrent flows leading to network events, such as congestion and microburst. To mitigate these network event effects, the XR system is required to adapt its behavior in injecting XR packets into the network. As the network here is very complex to be modeled (even though sometimes it can be modeled by network calculus in simplified scenarios), it is typically considered as a black-box. In this assumption, multiple agent reinforcement learning (RL) as shown in Fig.\ref{fig:MARL} can be a potentially very promising solution.

The objective here is to build a mapping relation (named a policy) from State to Action according to the Reward criterion. In order to be able to handle high-dimensional state space, a neural network is used to approximate this mapping relation. Once the network structure (a fully connected network in Fig.\ref{fig:MARL}) is defined, it is only needed to update the weight parameters of the neural network. Specifically, the learning strategy considered in this article is actor-critic where the actor fully explores the environment while respecting the previously learned policy most of the time (normally with $\epsilon$-greedy way). The critic uses the experience of a period from actors to calculate its policy and update its most up-to-date policy with actors regularly. It is noted that the temporal difference error is used here to efficiently update the policy by combining the expected reward and actually received reward through normalized weights. Detailed definitions are listed as follows.

\begin{itemize}
    \item State: Network measurements including throughput, latency, jitter, packet loss rate, etc.
    \item Action: Bitrate of the XR packets controlled by the XR headset end-host.
    \item Reward: A QoE metric composed of the factors discussed in Section \ref{Section:QoE}. An example of the reward function \cite{New_York_HoloLens} could be the combination of reward from the smoothness (i.e., the change of framerate), reward from the quality of the XR display (i.e., the change of resolution), and penalty from the latency.
    
    % \begin{equation}
    %     QoE(t)= \underbrace{\alpha\sum_{n=1}^N \log \frac{F_n}{F_{min}}}_\textrm{Smoothness}+\underbrace{\beta\sum_{n=1}^N \log \frac{R_n}{R_{min}}}_\textrm{Resolution Quality}-\underbrace{\gamma\sum_{n=1}^N \exp(\frac{L_n}{L_{min}})}_\textrm{Latency Penalty}
    % \end{equation}
    
    % \begin{align*}
    %     QoE(t)= & \underbrace{\alpha\sum_{n=1}^N \log \frac{F_n}{F_{min}}}_\textrm{Smoothness}+\underbrace{\beta\sum_{n=1}^N \log \frac{R_n}{R_{min}}}_\textrm{Resolution Quality} \\
    %     & -\underbrace{\gamma\sum_{n=1}^N \exp(\frac{L_n}{L_{min}})}_\textrm{Latency Penalty}
    % \end{align*}
    
    % where $F_n$ is average framerate, $R_n$ is average resolution, and $L_n$ is average latency within the time window of $n$ frames. $\alpha$, $\beta$, and $\gamma$ are hyperparameters to weigh the three factors. This QoE definition gives reward to the smoothness and quality of the XR display and penalizes the latency. 
    \item RL algorithm: If each RL agent is assumed responsible for one XR end-host device, multiple XR devices would require multiple RL agents working together. In this case, these agents share either learned policy, observations or experience with each other or their neighbors and achieve an overall high gain in reward. Off-the-shelf algorithms are with different features in continuous or discrete state/action space, and centralized or distributed style.
\end{itemize}

\subsection{XR-Aware Smart Network}
As mentioned in Section \ref{Section:Network Traffic Pattern}, the XR content is compressed before transmission. The basic idea of transmission based on MPEG compression is to transmit the keyframe and the temporal or spatial difference of the following a few frames. Specifically, the keyframe, sometimes called intra-frame or I frame for shot, is generated regularly. The predicted frame, P frame for short, describes the temporal or spatial difference between the current frame and its previous I frame or P frame. Bi-directional frame, B-frame for short, describes the difference of the current frame and its previous or future I and P frames. Thus, during the case of network congestion, the XR packets corresponding to different frames, mentioned as I, P, B frames, would be dropped in the network devices (e.g., switches or routers). If the dropped packets correspond to an I frame, the XR content is not able to be decompressed which would have a very severe effect in users' QoE. On the other hand, the P frame has the least effect on the final QoE which might be considered to be dropped when necessary. Once the XR packet shows the corresponding frame category in the packet header, the network could leverage the information in different levels of network management, such as queue management, load balancing, microburst mitigation, etc.

\section{Conclusion}
% \section{Metaverse Streaming Outlook}
% \hl{conclusion need to be added}
% \hl{Brief Future Outlook with Conclusion inside}
To enable the future promise of the Metaverse, it is therefore timely to seriously consider the high requirement of 360\textdegree{} stereoscopic streaming. In this article, simple calculations are used to clearly demonstrate the requirements in latency, bandwidth, and QoE. With the integration of the network and XR applications, a vision is provided of how technologies could evolve to support the Metaverse from a multimedia communication viewpoint.

% or


% \appendices

% \section{}

% \section*{Acknowledgment}


\ifCLASSOPTIONcaptionsoff
  \newpage
\fi



\bibliographystyle{IEEEtran}
\bibliography{refs}


% \end{thebibliography}


% \end{IEEEbiography}

\begin{IEEEbiographynophoto}{Haiwei Dong}
is a Principal Researcher with Huawei Technologies Canada, Ottawa, ON, Canada. His research interests include artificial intelligence, multimedia communication, multimedia computing, and robotics. Dong received his Ph.D. degree in Computer Science and Systems Engineering from Kobe University, Kobe, Japan. Contact him at haiwei.dong@ieee.org.
\end{IEEEbiographynophoto}

\vskip 0pt plus -1fil

\begin{IEEEbiographynophoto}{Jeannie S.A. Lee}
is an Associate Professor at the Singapore Institute of Technology. Her other appointments are the Director of Programmes and Deputy Director, Center for Immersification. She previously worked at Qualcomm on multimedia algorithms. She holds a Ph.D. in Computer and Electrical Engineering from the Georgia Institute of Technology. Contact her at jeannie.lee@singaporetech.edu.sg.
\end{IEEEbiographynophoto}

%\newpage



%\vfill


\end{document}